\documentclass[fleqn,square,numbers,sort&compress,reprint]{elsarticle}

\usepackage{bm}
\usepackage{amsmath}
\usepackage{amssymb}
\usepackage{revsymb}
\usepackage{subcaption}
\usepackage{graphicx}
\usepackage{fourier}
\usepackage{color}
\usepackage[colorlinks=true,linkcolor=blue,citecolor=blue,urlcolor=blue]{hyperref}
\usepackage[para]{manyfoot}
\usepackage{dsfont}

\DeclareMathAlphabet{\mathcal}{OMS}{cmsy}{m}{n}
\DeclareMathOperator{\sinc}{sinc}
\DeclareNewFootnote{B}[fnsymbol]
\DeclareSymbolFont{CMletters}{OML}{cmm}{m}{it}
\SetSymbolFont{CMletters}{bold}{OML}{cmm}{b}{it}
\DeclareMathSymbol{\ell}{\mathord}{CMletters}{"60}
\newcommand\BbbGamma{\reflectbox{\rotatebox[origin=c]{180}{$\mathds L$}}}
\begin{document}

\begin{frontmatter}
\title{A Novel Pseudo-Spectral Time-Domain Theory of Magnetic Neutron Scattering Illustrated Using A Uniformly Magnetized Sphere}

\author{Kun Chen}
\ead{kunchen@siom.ac.cn, kunchen@alum.mit.edu}
\affiliation{organization={Key Laboratory of Quantum Optics, Department of Aeorspace Laser Technology and Systems,
	Shanghai Institute of Optics and Fine Mechanics, Chinese Academy of Sciences},
   city={Shanghai},
   postcode={201800},
   country={China}}

\begin{abstract}
   A universal numerical method is developed for the investigation of magnetic
   neutron scattering.  By applying the pseudospectral-time-domain (PSTD) algorithm to
   the spinor version of the Schr\"{o}dinger equation, the evolution of the
   spin-state of the scattered wave can be solved in full space and
   time.  This extra spin degree of freedom brings some unique new features absent in the 
   numerical theory on the scalar wave scatterings~\cite{article:Chen2024a}.
   Different numerical stability condition has to be re-derived due to the
   coupling between the different spin components.  As the simplest application, the
   neutron scattering by the magnetic field of a uniformly magnetized sphere is
   studied.  The PSTD predictions are compared with those from the
   Born-approximation.  This work not only provides a systematic tool for
   analyzing spin-matter interactions, but also builds the forward model for
   testing novel neutron imaging methodologies, such as the newly developed
   thermal neutron Fourier-transform ghost imaging.
\end{abstract}


\end{frontmatter}


\section{Introduction}
Magnetic neutron scattering is the preferred method to probe materials' magnetic
structure.  As a neutral particle, neutron can penetrate deep into materials.
The nuclear magnetic moment of neutron leads to its direct interaction with
unpaired electrons of atoms.  Instrumental developments of various
experimental schemes have tremendously broadened the knowledge about vast
magnetic phenomena and disclosed the underlining
dynamics~\cite{book:Squires1978,book:Izyumov1970,book:Lovesey1984,book:YMZhu2005,book:Chatterji2006}.
Especially, the polarized neutron scattering has provided detailed and
unambiguous information about the magnetic structures.

Theoretical expressions for the spin-state cross-sections of magnetic neutron scattering
\textit{in the far field} have been derived from the Schr\"{o}dinger equation
using the
Fermi's golden rule~\cite{book:Squires1978}.  However, these mathematical formulas
of cross-sections mainly serve as a qualitative interpretation of the
experimental results.  The magnetic structure deduced from the data would be
represented by form factors in the theory.  Given a general magnetization
distribution, it is hard to quantitatively predict the scattering cross-section.
Furthermore, the existing theories only predict the far-field behavior.  When it
comes to the scattered waves in the near field and the mid field (i.e., the
Fresnel region), the traditional methods would be incapable to provide
reasonable results.

There is a separate motivation to calculate the magnetically scattered wave
function in the Fresnel region (FR).  Over the past two decades, there have been
intense developments on a lensless imaging methodology, referred to as the ghost
imaging (GI)~\cite{article:HSS-PRL2016,article:Paganin2016,article:Ratner2018,
article:Khakimov2016,article:kingston2020,article:WangKG-PRA2009,article:Chen2018}.
GI explores the fourth-order correlation of the wave fields.
In a typical GI setup, the spatially-incoherent incident beam, either a bosonic or a fermionic field,
is split into a sample arm and a reference arm.  Spatially resolved wavefront
detection is employed in the reference arm only, while the signals in the sample
arm are collected by a bucket detector.  The bosonic GI is enlightened by the
Hanbury Brown and Twiss (HBT) effect~\cite{HBT-Nature1956a,HBT-Nature1956b}, and
exploits the photon bunching phenomenon arising from the bosonic statistics of
optical fields~\cite{article:HSS-PRL2016,article:Paganin2016}.  On the other
hand, the fermionic GI, still at its early stage of development, utilizes the
antibunching phenomenon of fermionic matter
waves~\cite{article:WangKG-PRA2009,article:Chen2018}, dictated by the
Pauli's exclusive principle and observed in a series of
experiments~\cite{Klesel-nature2002,Rom-nature2006,Iannuzzi-PRL2006,Iannuzzi-PRA2011}.
The Fourier-transform (FT) implementation of both GIs can achieve microscopic
imaging of the matter, potentially with de Broglie wavelength level resolution.
Unlike the conventional imaging techniques which detect the far-field signals,
the FT GIs selectively detect the signals from the FR.  This is the
crucial difference between this new and the existing, old imaging strategies.

In recent years, we established a microscopic imaging theory for atomic and
magnetic structures based on thermal neutron FT GI.  The theory intends to
reconstruct the distribution of atom sites and material magnetizations from
experimental data.  This falls into the category of inverse problems.  To test the
validity and accuracy of this theory, a numerical simulation is necessary before
the actual experiment can be attempted.  In short, a forward model is needed to
generate the "experimental" data.  Neutron scattering by matter involves two
independent processes, nuclear scattering by the nuclei of atom sites and
magnetic scattering by the magnetic structure of the material.  Both belong to
the potential scattering problems of the Schr\"{o}dinger equation.
Because the range of nuclear forces is within 10 fm and the wavelengths of
thermal neutrons are at the order of 1 \r{A} to a few nm, the nuclear interaction
can be easily handled as a delta potential scattering~\cite{book:Squires1978}.  The actual difficulty
lies in the latter one, in which there is no easy way to calculate the
scattering by an arbitrary magnetic field.  In addition, fermionic
GI is a fully quantum exploitation in the FR.  This fact rules out the Monte
Carlo (MC) method.  MC treats the incidence as a bunch of particles undergoing a
series of collisions with a random matrix of scatterers.  The quantum behavior is only
reflected in the form of scattering phase function, i.e., the deflection of the particle's flying
direction after each collision is sampled in an approximate quantum way.  While between consecutive
collisions, the particle goes through straight line free-flying, ignoring all
wave nature and precluding any quantum cross-talks between paths.  In addition, the
scatterings at different sites are treated as independent ones, without any
interference.  Therefore, preserving the quantum behavior requires a startup
purely from the first principle, the Schr\"{o}dinger equation.  However,
conventional scattering analyses focus on the far field and thus are inadequate
for GI.  All these considerations demand a completely new treatment of the
magnetic scattering.

In the first step toward this final goal, we recently successfully built a
numerical theory to universally solve the quantum potential scattering wave
function in full 3D regions, ranging from the near field, to the FR, and to the
far field~\cite{article:Chen2024a}.  Yet this theory still lacks the spin degree
of freedom, and has to be extended to spinor wave functions before it can
applied to the magnetic studies.  In this paper, we will finish the extra steps
and present a forward model calculation on the neutron scattering by a magnetic
field.

\section{PSTD algorithm for solving the Schr\"{o}dinger equation of spinor wave}
The quantum scattering of the neutron wave by a magnetic field is governed
by the Schr\"{o}dinger equation~\cite{book:Squires1978},
\begin{equation}
   i\hbar\frac{\partial\Psi}{\partial t}=-\frac{\hbar^2}{2m_n}\nabla^2\Psi
   -\boldsymbol{\mu}_n\cdot\mathbf{B}\Psi. \label{eq:OriginalSchrodinger}
\end{equation}
Here, $m_n$ is the neutron mass, $\boldsymbol{\mu}_n=-\gamma\mu_N\boldsymbol{\sigma}$ the
magnetic dipole moment of the neutron, $\boldsymbol{\sigma}$ the Pauli spin operator
for the neutron, $\gamma=1.913$ and $\mu_N$ the nuclear magneton, $\mathbf{B}$ the
magnetic induction, and $\Psi$ the spinor wave function.  Apply the same
dimensionless procedure~\cite{article:Chen2024a}, i.e., define
\begin{equation}
   \tau=\omega_0
   t,\qquad (\bar{x},\bar{y},\bar{z})=(k_0x,k_0y,k_0z),
   \label{eq:ScaledCoordinates}
\end{equation}
where $\omega_0=E_0/\hbar$ and $k_0=P_0/\hbar=\sqrt{2m_nE_0}/\hbar$.  For a
monochromatic plane wave incidence, $E_0$ and $P_0$ would be the energy and
momentum of the neutron wave, respectively.  For a pulsed plane wave incidence,
$E_0$ and $P_0$ would be the values at the central frequency.  Using the
dimensionless variables 
of Eq.~(\ref{eq:ScaledCoordinates}), Eq.~(\ref{eq:OriginalSchrodinger}) is
rescaled to
\begin{equation}
   \frac{\partial\Psi}{\partial\tau}=i{\bar{\nabla}}^2\Psi-i\frac{\gamma\mu_N}
   {E_0}\boldsymbol{\sigma}\cdot\mathbf{B}\Psi,
   \label{eq:ScaledSchrodinger}
\end{equation}
and $\bar{\nabla}=\hat{x}\frac{\partial}{\partial\bar{x}}+\hat{y}\frac{\partial}
{\partial\bar{y}}+\hat{z}\frac{\partial}{\partial\bar{z}}$.  In this way, the rescaled
Schr\"{o}dinger equation becomes one on pure numbers.  This
greatly simplifies the derivation of numerical algorithms.

\begin{figure}
   \centering
   \includegraphics[width=0.75\linewidth]{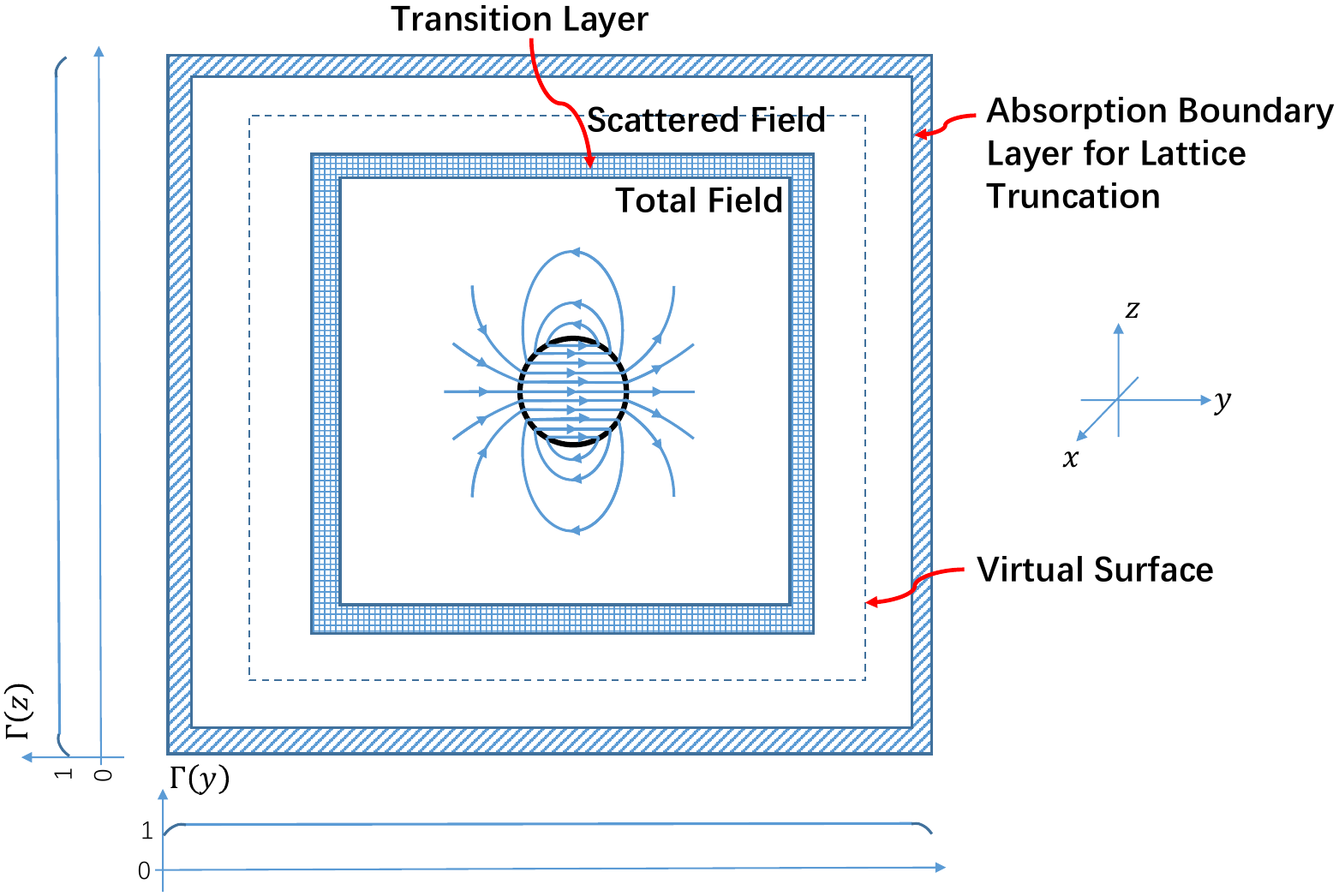}
   \caption{Setup of the Internal Model}
   \label{fig:mss-setup}
\end{figure}

There is no analytical solution to Eq.~(\ref{eq:ScaledSchrodinger}) for a
general distribution of magnetic field $\mathbf{B}$.  However, numerical
solution can be sought for a localized $\mathbf{B}$, even of an arbitrary shape.
Highly efficient parallel computing is inevitable.  The Pauli spin operators in
Eq.~(\ref{eq:ScaledSchrodinger}) are represented by $2\times 2$ matrices.  Thus
the spinor wave function $\Psi$ takes the form of a $2\times 1$ matrix,
containing two spin components.  The layout of the numerical lattice model here
is very similar to the scalar wave studies
(Fig.~\ref{fig:mss-setup})~\cite{article:Chen2024a}.  Again, we employ the
total-field/scattered-field (TF/SF) scheme and the PSTD approach to the partial
derivatives.  The terminology TF/SF stems from the computational
electrodynamics~\cite{book:Taflove2005}.  The term "field" in the
Schr\"{o}dinger world refers to the probing particle's matter wave field.  In
Fig.~\ref{fig:mss-setup}, the computation domain is divided into shell-like
regions.  The outmost numerical absorption boundary serves to truncate the space
lattice and converts the computation in infinite domain to finite domain.  Three
$\Gamma$ functions, one along each coordinate dimension, are multiplied to the
wave function at each iteration to impose the absorption boundary condition
(ABC).  The small smooth dips near the left and right ends of $\Gamma(y)$
attenuate the wave function near the left and right boundaries.  The boundaries in
the $x$ and $z$ directions are controlled in the same way by $\Gamma(x)$ and
$\Gamma(z)$.  The 3D ABC multiplier is the product 
$\BbbGamma(\mathbf{r})=\Gamma(x)\Gamma(y)\Gamma(z)$, and the 1D-$\Gamma$
function is chosen to be~\cite{article:Kosloff1986,article:Chen2024a}
\begin{equation}
   \Gamma(d)=\exp\left(-\frac{U_0}{\cosh^2(\alpha d)}\Delta\tau\right),
   \label{eq:Gamma}
\end{equation}
where $d$ is the distance (the smaller one) from the left or right boundaries.
The inner core in Fig.~\ref{fig:mss-setup}, referred to as the TF, encloses all
the magnetic fields and contains all the neutron-target interactions; whereas
the SF is the free-propagating region of the scattered waves.  In the TF, the
up/down spin components will inter-exchange due to the off-diagonal elements of
$\sigma_x$ and $\sigma_y$, whereas exterior to the TF, the up/down spin
components decouple and the spin states are preserved.  The transition layer
connects the TF and the SF, incorporates the incident wave into the model, and
smoothly converts the total wave function (in the TF) to the pure scattered
wave function (in the SF), based on the fact that
\begin{equation}
   \Psi^\text{total}=\Psi^\text{scat}+\Psi^\text{inc},\label{eq:psi_total_scat_inc}
\end{equation}
where $\Psi^\text{total}$, $\Psi^\text{scat}$ and $\Psi^\text{inc}$ are the
total, scattered and incident wave functions, respectively.  It is important to
stress that, unlike the electromagnetic studies where the TF could be tightened
to the permittivity and permeability structure and stop at the material
boundary, in the quantum case the magnetic field extends far beyond the underlying magnetization
structure.  Since the magnetic interaction is long-range, a reasonable cutoff on
range is required to limit the size of TF.  The field intensity of a magnetic
dipole typically decays as $r^{-3}$, so in practice a TF should be at least one
order of magnitude larger than the magnetization size.

Similar to the studies on the Maxwell's equations~\cite{article:GaoX2004} and
the Schr\"{o}dinger equation of scalar-waves~\cite{article:Chen2024a}, we introduce
a taper function
$\zeta(\mathbf{\bar{r}})=\zeta_x(\bar{x})\zeta_y(\bar{y})\zeta_z(\bar{z})$ on
the space lattice, such that $\zeta(\mathbf{\bar{r}})$ is 0 for
$\mathbf{\bar{r}}$ in the SF and ABC, 1 for $\mathbf{\bar{r}}$ in the TF, 
and smoothly rises from 0 to 1 for $\mathbf{\bar{r}}$ across the transition
layer.  The optimal choice of the $\zeta$
function is discussed in Ref.~\cite{article:Chen2024a}.
Consequently, a new wave function
\begin{equation}
   \tilde{\Psi}(\mathbf{\bar{r}},\tau)=\Psi^\text{scat}(\mathbf{\bar{r}},\tau)
   +\zeta(\mathbf{\bar{r}})\Psi^\text{inc}(\mathbf{\bar{r}},\tau)
   \label{eq:tildePsi}
\end{equation}
would represent the total wave function in the TF and the pure scattered wave function
in the SF.  Recall that both the incident wave function
$\Psi^\text{inc}(\mathbf{\bar{r}},\tau)$ and the scattered wave function
$\Psi^\text{sca}(\mathbf{\bar{r}},\tau)$ satisfy the Schr\"{o}dinger equation in free
space.  Combining with Eq.~(\ref{eq:ScaledSchrodinger}), we have
\begin{equation}
   \frac{\partial\tilde{\Psi}(\mathbf{\bar{r}},\tau)}{\partial\tau}
   =i{\bar{\nabla}}^2\tilde{\Psi}(\mathbf{\bar{r}},\tau)-i\frac{\gamma\mu_N}{E_0}
   \boldsymbol{\sigma}\cdot\mathbf{B}(\mathbf{\bar{r}})\tilde{\Psi}(\mathbf{\bar{r}},\tau)
   -i\left[{\bar{\nabla}}^2\zeta(\mathbf{\bar{r}})+2\bar{\nabla}\zeta(\mathbf{\bar{r}})
   \cdot\bar{\nabla}\right]\Psi^\text{inc}(\mathbf{\bar{r}},\tau).
   \label{eq:tildePsiSchrodinger}
\end{equation}
Because $\mathbf{B}(\mathbf{\bar{r}})=0$ for regions outside the TF, the
scattering potential is well defined in the entire computation domain.  As
$\tilde{\Psi}(\mathbf{\bar{r}},\tau)$ in the SF is automatically the scattered
wave function, the task of solving the original scattering problem in infinite
space is reduced to solving Eq.~(\ref{eq:tildePsiSchrodinger}) on the truncated
lattice of Fig.~\ref{fig:mss-setup}.

Now we explicitly express the spinor $\tilde{\Psi}$ in the matrix form
\begin{equation}
   \tilde{\Psi}=\binom{\psi_u}{\psi_d}.
\end{equation}
Substituting the Pauli matrices into Eq.~(\ref{eq:tildePsiSchrodinger}), we have
the explicit equations,
\begin{eqnarray}
   \frac{\partial\psi_u}{\partial\tau}&=&i{\bar{\nabla}}^2\psi_u-i\frac{\gamma\mu_N}{E_0}
   (B_z\psi_u+B_x\psi_d-iB_y\psi_d)\nonumber\\
   &&-i\left[{\bar{\nabla}}^2\zeta\psi^\text{inc}_u
   +2\bar{\nabla}\zeta\cdot\bar{\nabla}\psi^\text{inc}_u\right],\label{eq:spinorsSchrodinger1}\\
   \frac{\partial\psi_d}{\partial\tau}&=&i{\bar{\nabla}}^2\psi_d-i\frac{\gamma\mu_N}{E_0}
   (B_x\psi_u+iB_y\psi_u-B_z\psi_d)\nonumber\\
   &&-i\left[{\bar{\nabla}}^2\zeta\psi^\text{inc}_d
   +2\bar{\nabla}\zeta\cdot\bar{\nabla}\psi^\text{inc}_d\right].\label{eq:spinorsSchrodinger2}
\end{eqnarray}
The three terms on the r.h.s. of Eqs.~(\ref{eq:spinorsSchrodinger1}) and
(\ref{eq:spinorsSchrodinger2}) correspond to the kinetic energy, the potential
energy and the incident wave contribution, respectively.  Apply the central finite difference to
the time derivative on the l.h.s. and implement pseudospectral operation on the
kinetic energy term, we obtain the iteration updating formulas
\begin{eqnarray}
   \psi_u\Big|_{i,j,k}^{n+1}&=&\Gamma_{i,j,k}\left\{\psi_u
   \Big|_{i,j,k}^{n-1}+2\Delta\tau\left[K_u\Big|_{i,j,k}^n+
   P_u\Big|_{i,j,k}^n+I_u\Big|_{i,j,k}^n\right]\right\},
   \label{eq:updating1}\\
   \psi_d\Big|_{i,j,k}^{n+1}&=&\Gamma_{i,j,k}\left\{\psi_d
   \Big|_{i,j,k}^{n-1}+2\Delta\tau\left[K_d\Big|_{i,j,k}^n+
   P_d\Big|_{i,j,k}^n+I_d\Big|_{i,j,k}^n\right]\right\}
   \label{eq:updating2}
\end{eqnarray}
for indices $i,j,k$ over the entire lattice.  Here $\Gamma_{i,j,k}$ is the grid value
of the mask function $\BbbGamma(\mathbf{r})$, introduced as the ABC for the
truncation of the lattice, equivalent to a complex absorption potential
(CAP)~\cite{article:Kosloff1986,article:Silaev2018}.  The increments due to the
kinetic energy are
\begin{eqnarray}
   K_u\Big|_{i,j,k}^n&=&-i\eta\sum_{\ell=\bar{x},\bar{y},\bar{z}}
   \mathcal{F}_\ell^{-1}\left[k_\ell^2\mathcal{F}_\ell\left[\psi_u\right]\right]
   \Big|_{i,j,k}^n,\label{eq:K1}\\
   K_d\Big|_{i,j,k}^n&=&-i\eta\sum_{\ell=\bar{x},\bar{y},\bar{z}}
   \mathcal{F}_\ell^{-1}\left[k_\ell^2\mathcal{F}_\ell\left[\psi_d\right]\right]
   \Big|_{i,j,k}^n\label{eq:K2}
\end{eqnarray}
for indices $i,j,k$ over the entire lattice, and $\eta=\sinc(\Delta\tau)$ is a
correction factor to cancel the numerical phase velocity error caused by
the time discretization~\cite{article:Chen2024a}.  The increments due to the
potential energy are
\begin{eqnarray}
   P_u\Big|_{i,j,k}^n&=&-i\frac{\gamma\mu_N}{E_0}\left[B_z\Big|
   _{i,j,k}\psi_u\Big|_{i,j,k}^n+(B_x-iB_y)\Big|_{i,j,k}
   \psi_d\Big|_{i,j,k}^n\right],\label{eq:P1}\\
   P_d\Big|_{i,j,k}^n&=&-i\frac{\gamma\mu_N}{E_0}\left[(B_x+iB_y)
   \Big|_{i,j,k}\psi_u\Big|_{i,j,k}^n-B_z\Big|
   _{i,j,k}\psi_d\Big|_{i,j,k}^n\right]\label{eq:P2}
\end{eqnarray}
for indices $i,j,k$ within the TF zone.  The increments due to the injection of
the incident wave are
\begin{eqnarray}
   I_u\Big|_{i,j,k}^n&=&-i\left[{\bar{\nabla}}^2\zeta
   \psi^\text{inc}_u+2{\bar{\nabla}}\zeta\cdot{\bar{\nabla}}\psi^\text{inc}
   _u\right]\Big|_{i,j,k}^n,\label{eq:I1}\\
   I_d\Big|_{i,j,k}^n&=&-i\left[{\bar{\nabla}}^2\zeta
   \psi^\text{inc}_d+2{\bar{\nabla}}\zeta\cdot{\bar{\nabla}}\psi^\text{inc}
   _d\right]\Big|_{i,j,k}^n\label{eq:I2}
\end{eqnarray}
for indices $i,j,k$ within the transition layer only.  Note that the analytical
expressions of $\zeta$ and $\bar{\nabla}\zeta$ are known.  Under the incidence
of a monochromatic plane wave or a plane wave of Gaussian-shaped pulse, the
$\psi^\text{inc}$ and $\bar{\nabla}\psi^\text{inc}$ are also known analytically
beforehand.  Therefore, $I_u$ and $I_d$ can be prepared before the code run.  On
the other hand, for the plane wave incidence of a pulsed wavepacket of arbitrary
time-profile, as the wavepacket will spread over time during the propagation, a
separate, independent one-dimensional free-space Schr\"{o}dinger equation (i.e.,
with the potential equal to 0) is often solved concurrently with
Eqs.~(\ref{eq:spinorsSchrodinger1}) and (\ref{eq:spinorsSchrodinger2}) to provide
the $\psi_u^\text{inc}$, $\psi_d^\text{inc}$ and their first derivatives during
the time marching.

Because outside the TF the magnetic induction $\mathbf{B}=0$, the spin status of
the neutron wave is preserved.  The up and down components of the scattered
spinor wave act like two independent scalar waves.  The procedure on the
scattered scalar wave~\cite{article:Chen2024a} can be directly applied.  The
virtual surface data for the up and down waves are stored separately, and the
near-to-distant-field transformation can be performed independently for each
spin.

\section{Stability Condition}
The iterations (Eqs.~(\ref{eq:updating1})-(\ref{eq:updating2})) will diverge quickly if
the time-marching step is large than a bound.  However, the stability
condition for the spinor wave function (Eq.~(\ref{eq:OriginalSchrodinger})) cannot
employ the conclusion for the scalar wave function~\cite{article:Chen2024a}.  The
off-diagonal elements of $\boldsymbol{\sigma}\cdot\mathbf{B}$ create coupling terms
between the differential equations of the up and down components.  To proceed,
again we adopt the approach of Ref.~\cite{article:Soriano2004}.  The discrete form
of Eq.~(\ref{eq:OriginalSchrodinger}) is first separated into a temporal
eigenvalue problem and a spatial eigenvalue problem, i.e.,
\begin{eqnarray}
   i\hbar\frac{\Psi^{n+1}-\Psi^{n-1}}{2\Delta
   t}&=&\lambda_t\Psi^n,\label{eq:Stab_t}\\
   -\frac{\hbar^2}{2m}\nabla^2\Psi(\mathbf{r},t)+\gamma\mu_N\boldsymbol{\sigma}\cdot
   \mathbf{B}(\mathbf{r})\Psi(\mathbf{r},t)&=&\lambda_s\Psi(\mathbf{r},t).
   \label{eq:Stab_s}
\end{eqnarray}

The matrix $\boldsymbol{\sigma}\cdot\mathbf{B}(\mathbf{r})$ is Hermitian.  Its
eigenvalues are $\pm B(\mathbf{r})$, where
\begin{equation}
B(\mathbf{r})=\sqrt{B_x^2(\mathbf{r})+B_y^2(\mathbf{r})+B_z^2(\mathbf{r})}.
\end{equation}
Their corresponding eigenvectors (normalized) form the column vectors of a
transform $\mathcal{A}(\mathbf{r})$ such that
\begin{equation}
   \mathcal{A}^\dag(\mathbf{r})\mathcal{A}(\mathbf{r})=\mathcal{A}(\mathbf{r})
   \mathcal{A}^\dag(\mathbf{r})=I, \quad
   \mathcal{A}^\dag(\mathbf{r})\boldsymbol{\sigma}\cdot\mathbf{B}(\mathbf{r})
   \mathcal{A}(\mathbf{r})
   =\begin{pmatrix} B(\mathbf{r}) & 0 \\ 0 & -B(\mathbf{r}) \end{pmatrix}.
   \label{eq:AdagA}
\end{equation}

Because the transform $\mathcal{A}(\mathbf{r})$ is time irrelevant,\footnoteB{In case
the magnetic field contains a KHz-MHz sweeping content, for a thermal neutron of
~10meV, the cycle period of the de Broglie wave is at least 6 orders of magnitude
smaller than the sweep cycle.  The time variation of $\mathcal{A}^\dag$ is negligible
within $\Delta t$.  Thus, Eq.~(\ref{eq:Stab_At}) still remains valid.  The $B$ in
Eq.~(\ref{eq:OriginalStability}) takes the then-sweeping value, and $\Delta t$
can also follow the $B$ and be slowly swept over time.}
Eq.~(\ref{eq:Stab_t}) is equivalent to
\begin{equation}
   i\hbar\frac{\left[\mathcal{A}^\dag(\mathbf{r})\Psi\right]\Big|^{n+1}-\left[
      \mathcal{A}^\dag(\mathbf{r})\Psi\right]\Big|^{n-1}}{2\Delta t}
   =\lambda_t\left[\mathcal{A}^\dag(\mathbf{r})\Psi\right]\Big|^n,
   \label{eq:Stab_At}
\end{equation}
i.e., $\mathcal{A}^\dag(\mathbf{r})\Psi$ is a linear combination of the up and down components
of the original $\Psi$.  The derivation of the condition for $\lambda_t$ is the
same as the scalar wave function case~\cite{article:Chen2024a}, i.e.,
\begin{equation}
   \left|Re\left(\frac{\lambda_t\Delta t}{\hbar}\right)\right|\leqslant 1.
   \label{eq:lambda_t}
\end{equation}

In FFT language, the spatial eigenvalue problem Eq.~(\ref{eq:Stab_s}) converts to
\begin{equation}
   \lambda_s\left[\mathcal{A}^\dag(\mathbf{r})\Psi\right]=\frac{\hbar^2}{2m}
   \mathcal{A}^\dag(\mathbf{r})\sum_{\ell=x,y,z}\mathcal{F}_\ell^{-1}\left[k_\ell^2\mathcal{F}_\ell
   \left[\Psi\right]\right]+\gamma\mu_N
   \begin{pmatrix} B(\mathbf{r}) & 0 \\ 0 & -B(\mathbf{r})\end{pmatrix}
   \left[\mathcal{A}^\dag(\mathbf{r})\Psi\right],\label{eq:Stab_As}
\end{equation}
where $\mathcal{F}_\ell$ and $\mathcal{F}_\ell^{-1}$ are the 1D FFT and its inverse
along the $\ell$ ($\ell=x,y,z$) direction.  Again we can apply the fact that the
kinetic energy representable by a PSTD lattice has an upper limit.  The maximum
$k$s on the r.h.s. of Eq.~(\ref{eq:Stab_As}) are $\pm\pi/\Delta x$,
$\pm\pi/\Delta y$ and $\pm\pi/\Delta z$ respectively.  Any larger $k$ will be
aliased to a value within the limits.  When we replace the $k_x$, $k_y$ and $k_z$
with these maximum values, they can be extracted out from the inverse FFTs,
resulting in
$\mathcal{F}_\ell^{-1}\left[\mathcal{F}_\ell\left[\Psi\right]\right]=\Psi$,
($\ell=x,y,z$).  Therefore the kinetic energy term is bounded below
\begin{equation}
   \frac{\hbar^2\pi^2}{2m}\left(\frac{1}{\Delta x^2}+\frac{1}{\Delta y^2}
   +\frac{1}{\Delta z^2}\right)\left[\mathcal{A}^\dag(\mathbf{r})\Psi\right].
\end{equation}
We now notice the transformed wave function $\mathcal{A}^\dag(\mathbf{r})\Psi$ has a
diagonalized upper bound form, and the upper bound for $\lambda_s$ can be
determined as
\begin{equation}
   \lambda_s < \frac{\hbar^2\pi^2}{2m}\left(\frac{1}{\Delta x^2}+\frac{1}{\Delta
   y^2}+\frac{1}{\Delta z^2}\right)+\gamma\mu_N B_\text{max}.\label{eq:lambda_s}
\end{equation}
Finally, combining Eqs.~(\ref{eq:lambda_t}) and (\ref{eq:lambda_s}), the
consistency requirement $\lambda_t=\lambda_s$ leads to the sufficient(but not
the necessary) stability condition of PSTD for the spinor wave function
\begin{equation}
   \Delta t\leqslant\frac{\hbar}{\frac{\hbar^2\pi^2}{2m}\left[\frac{1}{\Delta
   x^2}+\frac{1}{\Delta y^2}+\frac{1}{\Delta z^2}\right]+\gamma\mu_N
   B_\text{max}}. \label{eq:OriginalStability}
\end{equation}
The dimensionless version of Eq.~(\ref{eq:OriginalStability}) is 
\begin{equation}
   \Delta\tau\leqslant\left[\pi^2\left(\frac{1}{\Delta\bar{x}^2}+\frac{1}
   {\Delta\bar{y}^2}+\frac{1}{\Delta\bar{z}^2}\right)+\frac{\gamma\mu_N
   B_\text{max}}{E_0}\right]^{-1}. \label{eq:ScaledStability}
\end{equation}

\section{Born approximation}
Let the incident plane wave function be
$\Psi^\text{inc}={\displaystyle\binom{s_u}{s_d}}e^{i\mathbf{k}\cdot\mathbf{r}}$ with
$\left|s_u\right|^2+\left|s_d\right|^2=1$.  The Born approximation to
Eq.~(\ref{eq:OriginalSchrodinger}) leads to the approximated solution of the
spinor wave function
\begin{equation}
   \Psi=\binom{s_u}{s_d}e^{i\mathbf{k}\cdot\mathbf{r}}-\frac{\gamma
   m_n\mu_N}{2\pi\hbar^2}\frac{e^{ikr}}{r}
   \int d^3r^\prime e^{-i\mathbf{q}\cdot\mathbf{r^\prime}}\begin{pmatrix}
   B_z(\mathbf{r^\prime}) & B_x(\mathbf{r^\prime})-iB_y(\mathbf{r^\prime})\\
   B_x(\mathbf{r^\prime})+iB_y(\mathbf{r^\prime}) &
   -B_z(\mathbf{r^\prime})\end{pmatrix}\binom{s_u}{s_d},\label{eq:SchrodingerNeutronBorn}
\end{equation}
with $\mathbf{q}=\mathbf{k^\prime}-\mathbf{k}$ the transferred wave vector
and $\mathbf{k^\prime}=k\,\mathbf{r}/r$ the wave vector of the outgoing spherical
wave.

\section{Modeling the neutron scattering by a uniformly magnetized sphere}
A uniformly magnetized sphere is the simplest, realistic
magnetic model.  Its magnetic field carries a cylindrical symmetry around the
internal magnetization.  The magnetic induction $B$ can be analytically
derived and behaves as a magnetic dipole in the far field~\cite{book:Jackson1999}.
Furthermore, the volume integration in the scattering amplitude term on the r.h.s
of Eq.~(\ref{eq:SchrodingerNeutronBorn}) can be analytically carried
out~(\ref{sec:appendix}), avoiding the uncertainties coming from numerical
integrations.  A comparison between the PSTD predictions and the Born
approximation can easily demonstrate the advantages of our method and quantify
the limits of the latter.

Figure~\ref{fig:mss-setup} employs a sphere of radius $a$ with a uniform magnetization
$\mathbf{M}_0$ along the $y$-axis.  The magnetic induction can be derived from
the magnetic scalar potential~\cite{book:Jackson1999}, leading to
\begin{equation}
   \mathbf{B}_\infty(\mathbf{r})=
   \begin{cases}
      \dfrac{2}{3}\mu_0 M_0 \hat{y}, & \text{if } r<a\\
      \mu_0
      M_0\dfrac{a^3}{r^3}\left[\dfrac{xy}{r^2}\hat{x}+\left(\dfrac{y^2}{r^2}
      -\dfrac{1}{3}\right)\hat{y}+\dfrac{yz}{r^2}\hat{z}\right], & \text{if } r\ge a
   \end{cases}
   \label{eq:umsB}
\end{equation}
An overall coefficient
\begin{equation}
   V_0=\gamma \mu_N\mu_0 M_0,\label{eq:V0}
\end{equation}
characterizes the energy-scale of the interacting potential and carries the
dimension unit of energy.  In the situation of monochromatic plane wave incidence,
Eqs.~(\ref{eq:spinorsSchrodinger1}) and (\ref{eq:spinorsSchrodinger2}) only
depend on two ratios, $V_0/E_0$ and $a/\lambdabar_0$.  A scaling law of the
differential scattering cross-section immediately follows, i.e.,
\begin{equation}
   \frac{d\sigma}{d\Omega}\left(E_0, V_0, a\right)=\lambdabar_0{}^2
   \frac{d\sigma}{d\Omega}\left(\frac{V_0}{E_0},\frac{a}{\lambdabar_0}\right),
   \label{eq:cross-section-scaling}
\end{equation}
and consequently the absolute values of $E_0$, $V_0$, $a$ no longer matter, and
the only length scale $\lambdabar_0$ is factorized out.  The
$\frac{d\sigma}{d\Omega}$ on the
r.h.s. of Eq.~(\ref{eq:cross-section-scaling}) depends on the two ratios and is 
dimensionless.

Outside the sphere, $\mathbf{B}_\infty$ decays as $r^{-3}$. A cutoff of the interaction
range is necessary for PSTD to proceed on a finite lattice.  We set the
$r$-cutoff to $b=8.175a$, corresponding to 3-order of magnitude decay of
interaction.  To be exact, the actual magnetic induction under study is
\begin{equation}
   \mathbf{B}_b(\mathbf{r})=\mathbf{B}_\infty(\mathbf{r})\,\Theta(r-b),
   \label{eq:umsBb}
\end{equation}
with $\Theta(x)$ a step function, such that when $x<0$, $\Theta(x)=1$, and
$x>0$, $\Theta(x)=0$.

Polarized magnetic neutron scattering exhibits anisotropic patterns.  The
scattering plane is best represented by its Euler angles.  We use the
y-convention for $(\alpha,\beta,\gamma)$~\cite{book:Sakurai2011}.  The first two
Euler angles for two characteristic planes are: the $x-y$ plane, 
$(\alpha,\beta)=(0^\circ,0^\circ)$; and the $y-z$ plane,
$(\alpha,\beta)=(0^\circ,90^\circ)$.  The $\gamma$ angle would easily label the
direction of the outgoing wave on the plane.

To obtain the following results, we select a magnetic sphere of radius equal to
one neutron wavelength, i.e., $a/\lambdabar_0=2\pi$.  The neutron plane wave is
incident along the $y$-axis.  An overlapping domain decomposition of $1\times
2\times 1$ topology is mapped to a two-node parallel platform.  A lattice of
$512\times 320\times 512$ grids is created on each subdomain, with
$\Delta\bar{x}=\Delta\bar{y}=\Delta\bar{z}=\pi/10$, corresponding to 20 grids
per wavelength.  The time increment is $\Delta\tau=\pi/1000$, at the requirement
of Eq.~(\ref{eq:ScaledStability}).  The widths of the ABC, the SF, the transition
layer, and the overlapping halo are 40, 41, 12 and 15 grids, respectively.  The
parameters for the ABC multiplier (Eq.~(\ref{eq:Gamma})) are
$U_0=5.0$ and $\alpha=0.1/\text{grid}$.  The virtual surfaces are set right on
the six middle planes of the SF.

To facilitate the comparison with the PSTD numerical results, we perform the
volume integral on the r.h.s. of Eq.~(\ref{eq:SchrodingerNeutronBorn}) upto the
cutoff radius $r^\prime=b$ (Eq.~(\ref{eq:BornPsiScat-b})).  A full Born
approximation is straightforward by taken $b\to\infty$.

\begin{figure}
   \centering
   \begin{subfigure}{0.45\linewidth}
      \includegraphics[width=\linewidth]{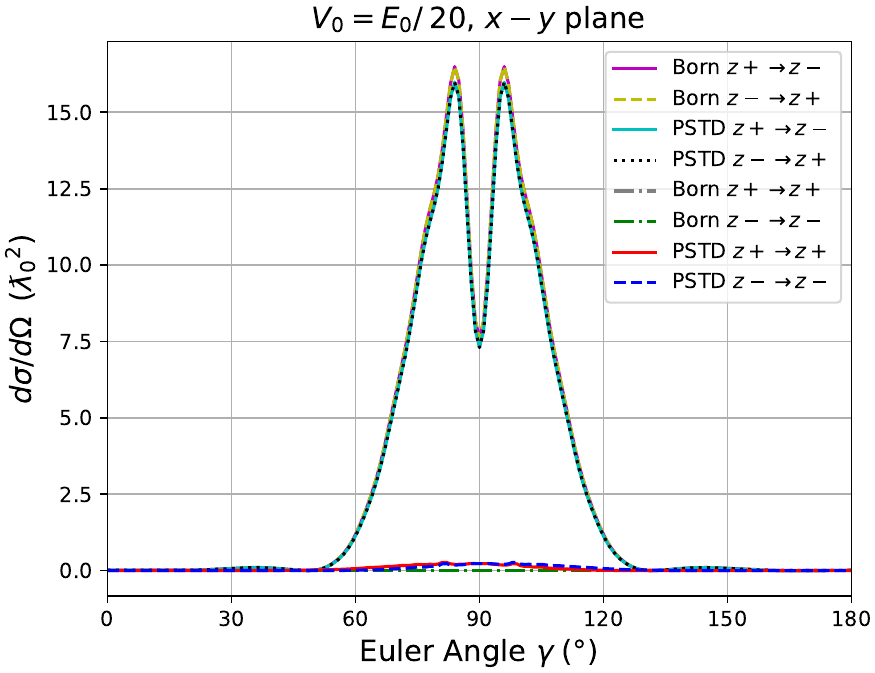}
      \label{fig:E20_x-y}
   \end{subfigure}
   \hfill
   \begin{subfigure}{0.45\linewidth}
      \includegraphics[width=\linewidth]{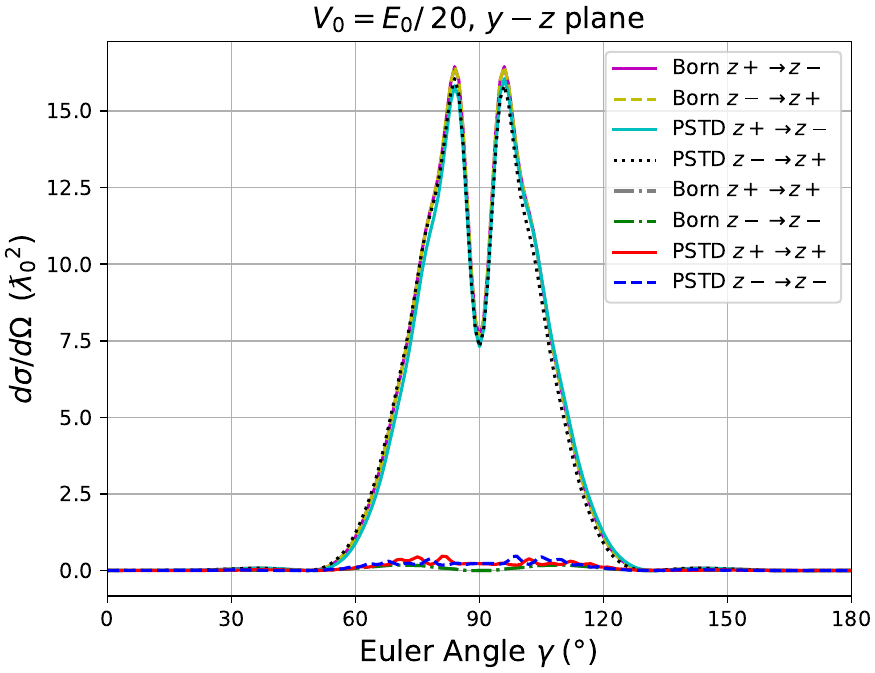}
      \label{fig:E20_y-z}
   \end{subfigure}
   \caption{Differential cross-sections of the neutron scattering by extra weak
   magnetic field $V_0=E_0/20$. The cutoff is at $b=8.175a$.}
   \label{fig:E20}
\end{figure}

\begin{figure}
   \centering
   \begin{subfigure}{0.45\linewidth}
      \includegraphics[width=\linewidth]{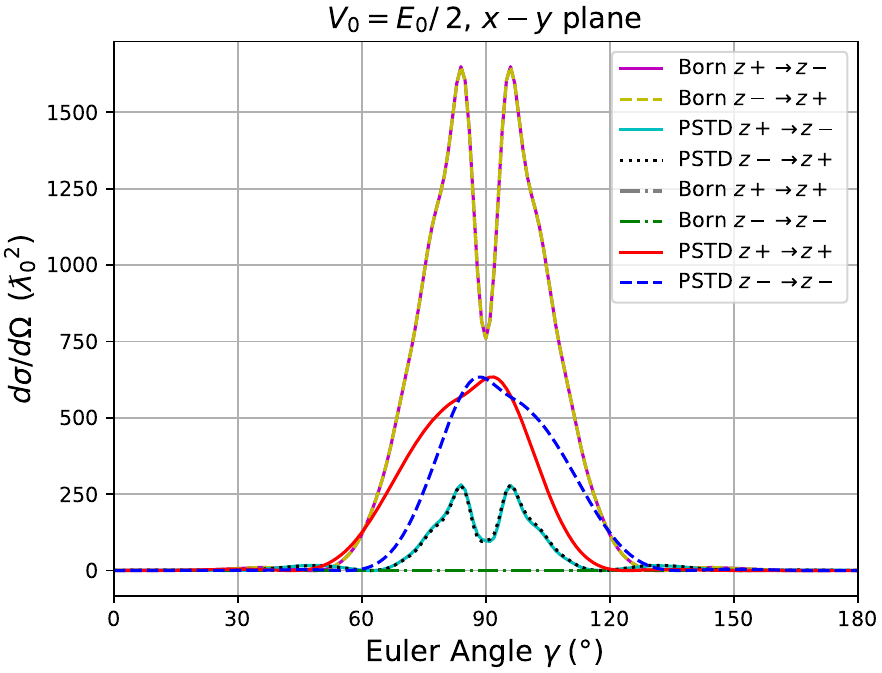}
      \label{fig:E2_x-y}
   \end{subfigure}
   \hfill
   \begin{subfigure}{0.45\linewidth}
      \includegraphics[width=\linewidth]{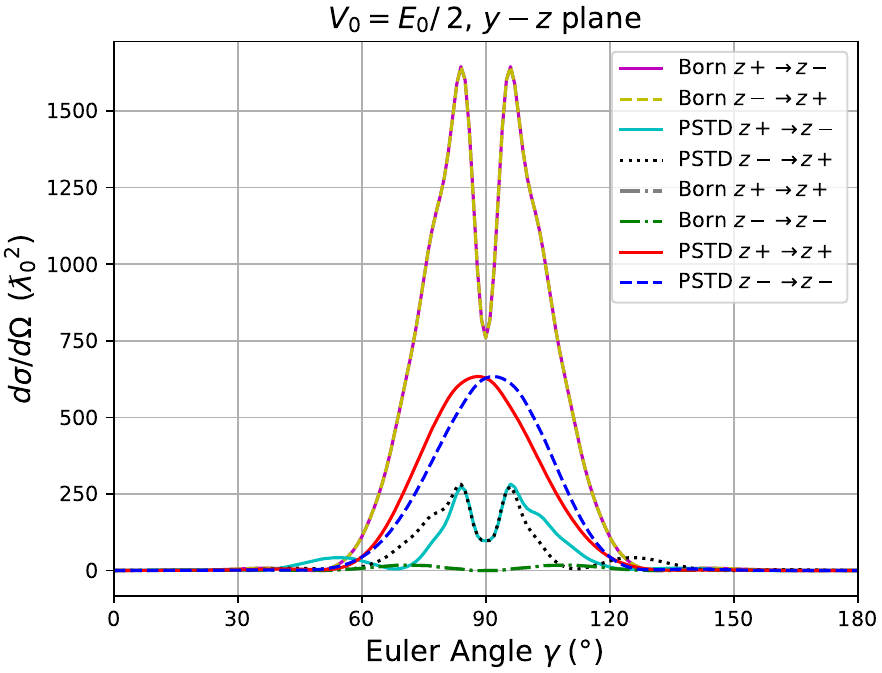}
      \label{fig:E2_y-z}
   \end{subfigure}
   \caption{Differential cross-sections of the neutron scattering by weak
   magnetic field $V_0=E_0/2$. The cutoff is at $b=8.175a$.}
   \label{fig:E2}
\end{figure}

\begin{figure}
   \centering
   \begin{subfigure}{0.45\linewidth}
      \includegraphics[width=\linewidth]{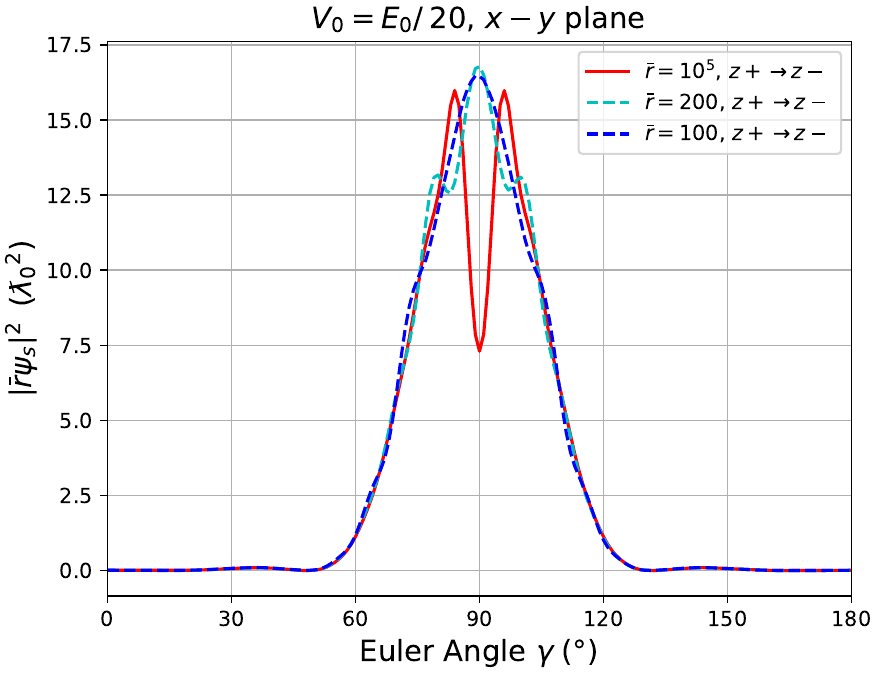}
      \label{fig:Far-Fresnel_E20_x-y}
   \end{subfigure}
   \hfill
   \begin{subfigure}{0.45\linewidth}
      \includegraphics[width=\linewidth]{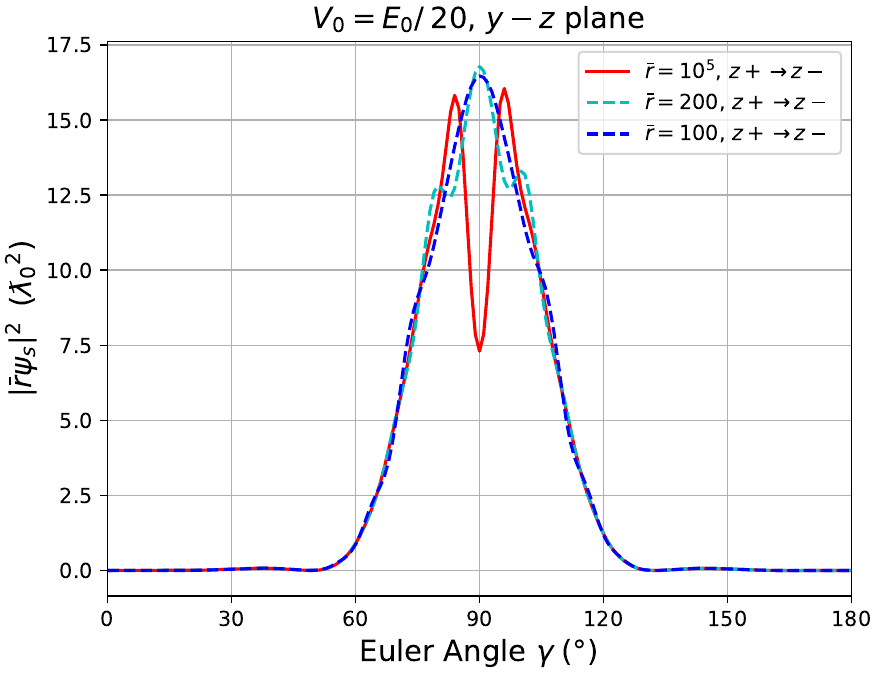}
      \label{fig:Far-Fresnel_E20_y-z}
   \end{subfigure}
   \caption{The PSTD solutions to the $z^+\to z^-$ scattering in the FR
   ($\bar{r}=100$ and $\bar{r}=200$) vs\@.  in the far-field ($\bar{r}=10^5$).
   The magnetic interaction strength $V_0=E_0/20$. The cutoff is at $b=8.175a$.}
   \label{fig:Far-Fresnel_E20}
\end{figure}

\begin{figure}
   \centering
   \begin{subfigure}{0.45\linewidth}
      \includegraphics[width=\linewidth]{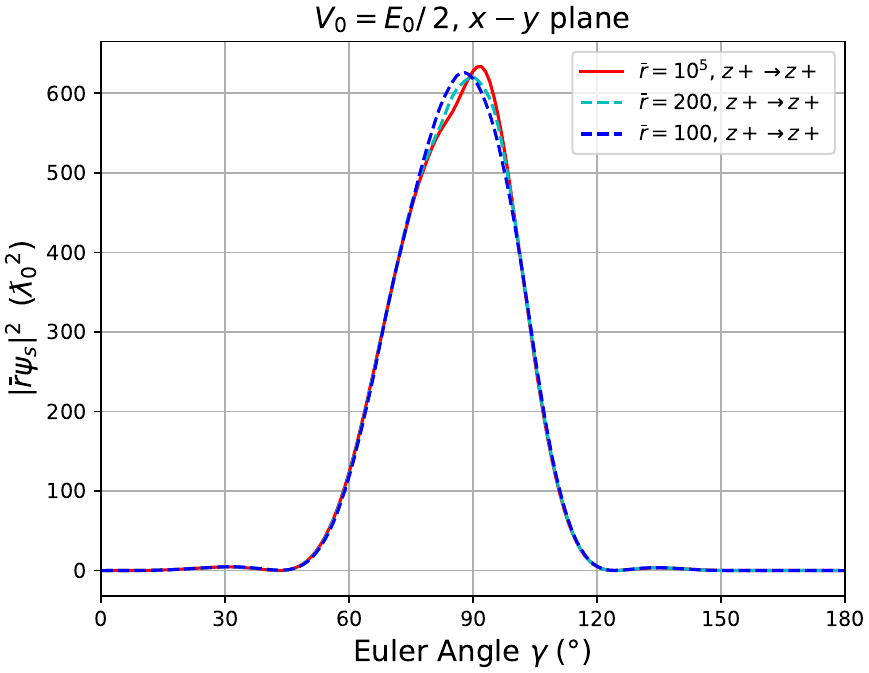}
      \label{fig:Far-Fresnel_E2_x-y}
   \end{subfigure}
   \hfill
   \begin{subfigure}{0.45\linewidth}
      \includegraphics[width=\linewidth]{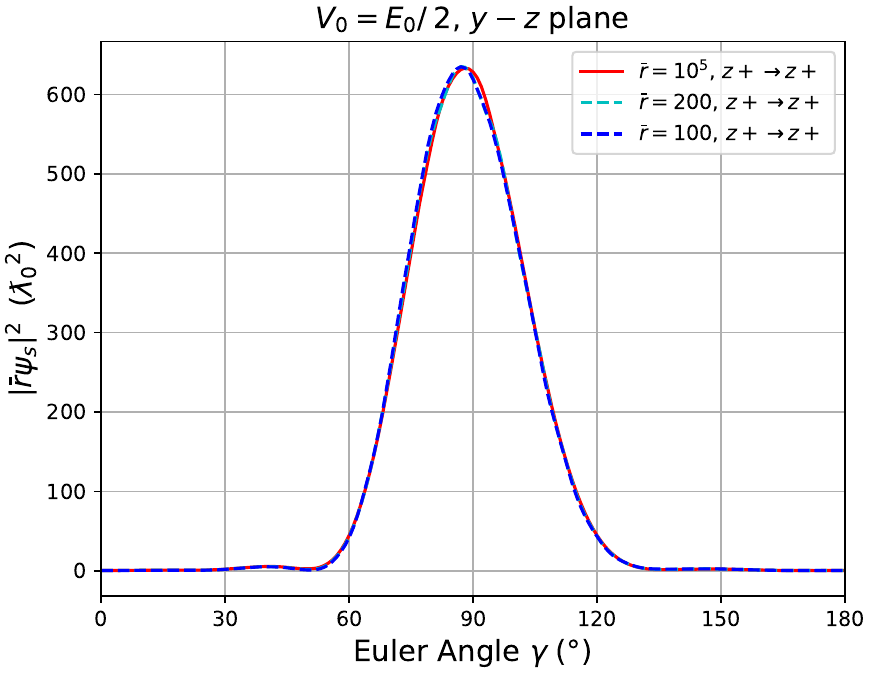}
      \label{fig:Far-Fresnel_E2_y-z}
   \end{subfigure}
   \caption{The PSTD solutions to the $z^+\to z^+$ scattering in the FR
   ($\bar{r}=100$ and $\bar{r}=200$) vs\@.  in the far-field ($\bar{r}=10^5$).
   The magnetic interaction strength $V_0=E_0/2$. The cutoff is at $b=8.175a$.}
   \label{fig:Far-Fresnel_E2}
\end{figure}

\begin{figure}
   \centering
   \begin{subfigure}{0.45\linewidth}
      \includegraphics[width=\linewidth]{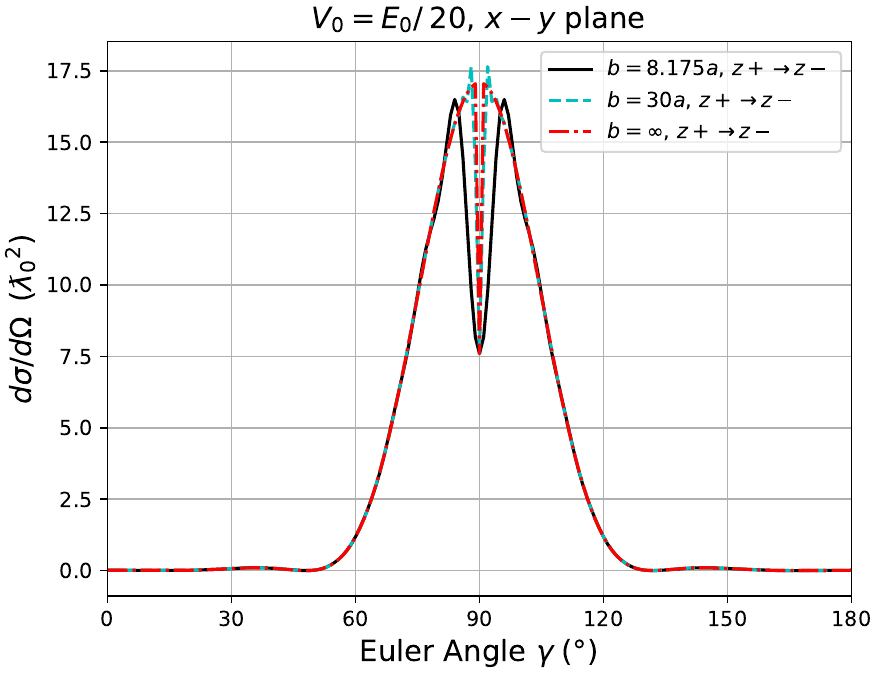}
      \label{fig:Born_CutoffFull_x-y}
   \end{subfigure}
   \hfill
   \begin{subfigure}{0.45\linewidth}
      \includegraphics[width=\linewidth]{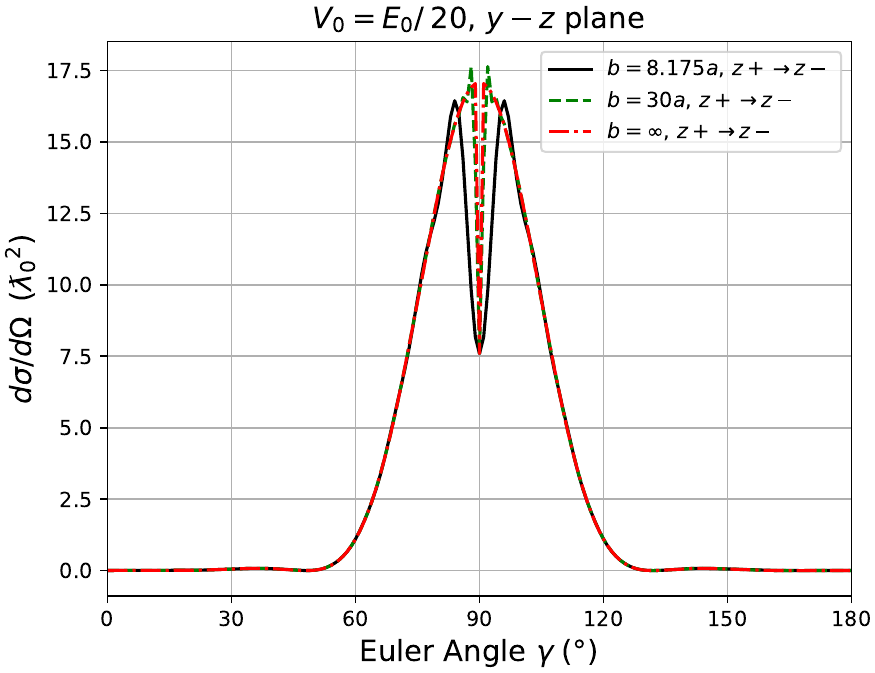}
      \label{fig:Born_CutoffFull_y-z}
   \end{subfigure}
   \caption{Effect of the interaction range cutoff on the results given by the Born approximation.
   The magnetic interaction strength $V_0=E_0/20$.}
   \label{fig:Born_CutoffFull}
\end{figure}

The simulations are conducted for two interaction strengths, $V_0=E_0/20$ and
$V_0=E_0/2$.  Both the PSTD and the Born-approximation results are shown
together on Figs.~\ref{fig:E20} and \ref{fig:E2}.  On the left panels, the Euler
$\gamma$ angle sweeps from the $+x$-axis through the $+y$-axis and to the
$-x$-axis, while on the right panels, $\gamma$ sweeps from the $-z$-axis through
the $+y$-axis and to the $+z$-axis.  The interaction-strength dependent
behaviors are not only manifested in the shape of the differential
cross-sections, but also in the spin state to spin state transitions.  For
example, when the interaction is weak $V_0=E_0/20$ (Fig.~\ref{fig:E20}), the
$|z+\rangle\to|z-\rangle$ has higher probability than the
$|z+\rangle\to|z+\rangle$, while it is the opposite when the interaction is 10
times as strong (Fig.~\ref{fig:E2}).

In the Born-approximation, the interaction strength $V_0$ enters as an overall
factor in the scattered amplitude.  When $V_0$ goes from
$E_0/20$ to $E_0/2$, the shape of its differential cross-section remains the
same, but the value becomes 100 times as large.

As expected, when the interaction is weak, the Born-approximation is very close
to the exact result of PSTD~(Fig.~\ref{fig:E20}).  However, when the potential
becomes stronger (Fig.~\ref{fig:E2}), the Born-approximation breaks down
quickly.  Thus, the PSTD solution can quantitatively assess the validity of the
Born-approximation.

There exists an axial symmetry in this specific example we present.  The
magnetization $\mathbf{M}_0$ is aligned on the $y$-axis and the incident wave is
also along the $y$-direction.  A rotation of the $|z+\rangle$ spin state
around the $y$-axis by $180^\circ$ would give the $|z-\rangle$ spin state,
and vice versa.  Thus 
Figs.~\ref{fig:E20}-\ref{fig:Far-Fresnel_E2} exhibit a flip-symmetry between
the $z+\to z+$ and $z-\to z-$ scatterings, as well as between the $z+\to z-$ and
$z-\to z+$ scatterings.  Another feature in Fig.~\ref{fig:E2} is that the
cross-sections show different shapes on the $x-y$ plane and on the $y-z$ plane.

Further numerical investigations using other incident spins reveal that the PSTD
algorithm automatically obeys the supposition principle of quantum states, for example
$|x+\rangle=\frac{1}{\sqrt{2}}|z+\rangle+\frac{1}{\sqrt{2}}|z-\rangle$.
Therefore, the four scattering wave functions ($z+\to z+$, $z+\to z-$, $z-\to
z+$, and $z-\to z-$) form a complete set of the spin degree of freedom.  The scattering
wave function of arbitrary incident spin and outgoing spin can be expressed as a linear
combination of the four basis.

Neutron ghost imaging detects the scattered neutrons in the FR.  The magnetized
sphere in this example is $2\lambda_0$ in diameter.  Its effective field spans
$~20\lambda_0$.  A distance $r\approx 400\lambda_0$ (i.e., $\bar{r}=800\pi$) marks
the transition from the FR to the far-field.  In Figs.~\ref{fig:Far-Fresnel_E20}
and \ref{fig:Far-Fresnel_E2} we present the PSTD results at three distances,
$\bar{r}=100$, $\bar{r}=200$, and $\bar{r}=10^5$.  Big differences mainly exist
in the forward angles, indicating the outgoing wave contains non-spherical wave
content.  This content evolves as the wave propagates.

\emph{Effect from the cutoff of the interaction range}\ \ The magnetic induction
$\mathbf{B}$ of a magnetic dipole decays as $r^{-3}$.  A numerical model, such as
the PSTD we present, must constrain the lattice size.  A cutoff on the
interaction range is unavoidable.  So far in this work, all the results are
obtained under the cutoff $b=8.175a$, with $a$ the sphere radius.  What will
change if we push $b$ beyond this limit?  In Fig.~\ref{fig:E20}, we see the
results of the PSTD and the Born-approximation agree very well when the
interaction is 5\% the neutron energy.  Under this limit, the Born-approximation
can provide qualitative assessment of this cutoff.  Using the analytical
expression Eq.~(\ref{eq:BornPsiScat-b}), the differential cross-sections are
plotted for $b=8.175a$, $b=30a$ and $b\to\infty$
(Fig.~\ref{fig:Born_CutoffFull}).  We notice the differences are localized in
the forward direction.  In scattering experiments (as opposite to transmission
ones), the forward angles are normally blocked by beam stop to absorb the
unscattered incident neutrons.  For $b=8.175a$, this discrepancy occurs within
$\gamma\approx 90^\circ\pm 6^\circ$.  An increase of $b$ narrows the angle
span.

\section{Summary}
In this work, we incorporate the spin degree of freedom into the PSTD solver to
the quantum potential scattering problems.  Because the different spin
components are coupled through the off-diagonal elements of the interaction
matrix, the PSTD algorithm for the scalar Schr\"{o}dinger equation cannot be
directly applied.  The extension to the spinor version requires the simultaneous
time-marching of both spin components satisfy a new stability condition.  Once
the virtual surface data of the up and down spin states are collected
separately, the scalar version of the near-to-distant-field transform can be
directly applied independently to each spin data.  This lies on the fact that the
magnetic field is absent outside the virtual enclosure, and thus the neutron
spin will not precess while propagating out.  The virtual surface data contain
the complete information of the spinor wave function at any distant locations.
Based on this development, magnetic neutron scattering can now be quantitatively
investigated.  In this paper, a uniformly magnetized sphere is used as an
example to demonstrate the capability and accuracy of the algorithm.  Much more
complicated magnetic structure can be constructed and its neutron scattering
straightforwardly calculated.   This forms the forward model of the newly
developed neutron ghost imaging (NGI) methodology~\cite{article:Chen2018}.  By
feeding the forward data to the NGI, the magnetization distribution can be
reconstructed and compared with the original setup.  Thus, the accuracy and
efficacy of NGI can be evaluated.  At last, our work supplies a numerical tool for
other neutron magnetic analysis technologies as well.

\section*{Acknowledgements}
This work was supported by the National Natural Science Foundation of
China under Grant Project No. 12075305.

\appendix
\section{The volume integral involed in the Born-approximation\label{sec:appendix}}
\renewcommand{\theequation}{A.\arabic{equation}}
\setcounter{equation}{0}
\setcounter{figure}{0}
In the following detailed theoretical calculation on the Born approximation, it
is the most convenient to adopt the conventional notation of the coordinate
system (Fig.~\ref{fig:appendix}), i.e., the direction of the magnetization
$\mathbf{M}_0$ is taken as the $z$-axis.  For consistency, at the last step when
Figs.~\ref{fig:E20} and \ref{fig:E2} are ploted, the $(\theta_q,\phi_q)$ angles in
Fig.~\ref{fig:appendix} are translated into the Euler angles in
Fig.~\ref{fig:mss-setup}.  In addition, the spin states are also rotated using
the D matrix.

\begin{figure}
   \centering
   \includegraphics[width=0.5\linewidth]{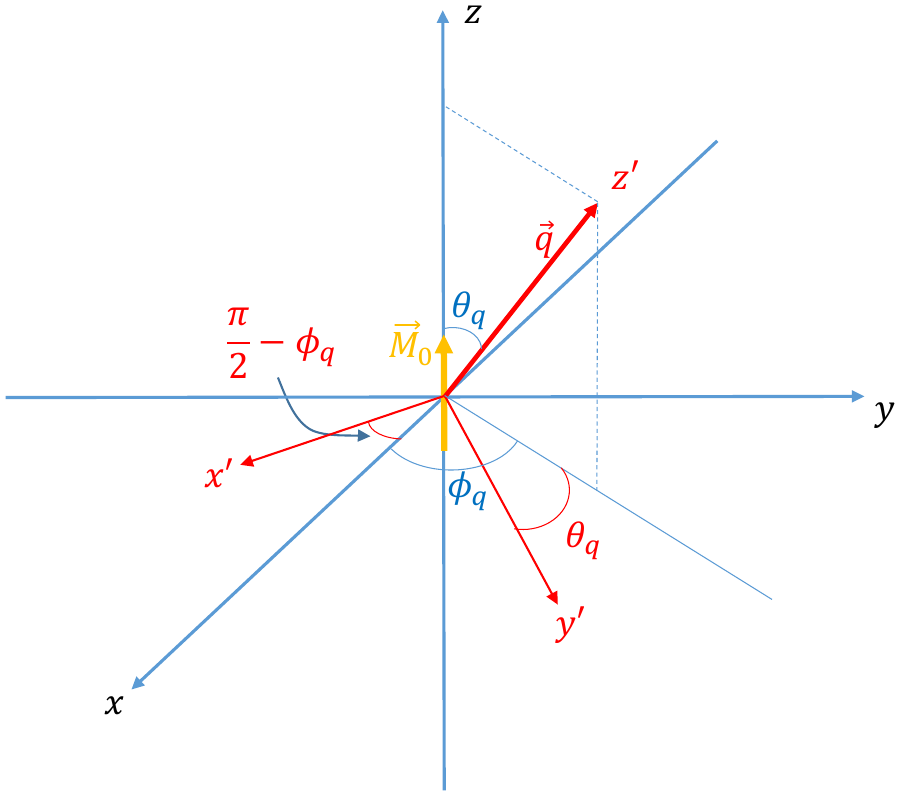}
   \caption{The $x-y-z$ and $x^\prime-y^\prime-z^\prime$ coordinate systems and
   the definition of the angles for the Born approximation calculation.}
   \label{fig:appendix}
\end{figure}

With the arrangement in Fig.~\ref{fig:appendix}, the magnetic induction becomes
\begin{equation}
   \mathbf{B}_\infty(\mathbf{r})=
   \begin{cases}
      \dfrac{2}{3}\mu_0 M_0 \hat{z}, & \text{if } r<a\\
      \mu_0
      M_0\dfrac{a^3}{r^3}\left[\dfrac{xz}{r^2}\hat{x}+\dfrac{yz}{r^2}\hat{y}
      +\left(\dfrac{z^2}{r^2}-\dfrac{1}{3}\right)\hat{z}\right], & \text{if } r\ge a
   \end{cases}
   \label{eq:appendix_umsB2}
\end{equation}
The scattering amplitude from the Born approximation of
Eq.~(\ref{eq:SchrodingerNeutronBorn}) is essentially the linear combination of three
volume integrals, i.e.,
\begin{equation}
   \mathcal{I}_\ell=-\frac{\gamma\mu_N m_n}{2\pi\hbar^2}\int d^3r\,e^{-i\mathbf{q}\cdot\mathbf{r}}
   B_\ell(\mathbf{r}),\quad \ell=x,y,z.
   \label{eq:appendix_Bxyz}
\end{equation}

Because $E_0=\hbar^2/(2m_n\lambdabar_0{}^2)$, the neutron mass $m_n$ in
Eq.~(\ref{eq:appendix_Bxyz}) can be expressed in terms of neutron energy.  Substituting
Eq.~(\ref{eq:appendix_umsB2}) into Eq.~(\ref{eq:appendix_Bxyz}) gives
\begin{eqnarray}
   \mathcal{I}_x&=&-\frac{V_0}{4\pi E_0\lambdabar_0{}^2}\int\limits_{a<r\leqslant
   b}d^3r\,\frac{a^3xz}{r^5}e^{-i\mathbf{q}\cdot\mathbf{r}},\label{eq:appendix_Ix}\\
   \mathcal{I}_y&=&-\frac{V_0}{4\pi E_0\lambdabar_0{}^2}\int\limits_{a<r\leqslant
   b}d^3r\,\frac{a^3yz}{r^5}e^{-i\mathbf{q}\cdot\mathbf{r}},\label{eq:appendix_Iy}\\
   \mathcal{I}_z&=&-\frac{V_0}{4\pi E_0\lambdabar_0{}^2}\left[\frac{2}{3}\int\limits_{r\leqslant
   a}d^3r\,e^{-i\mathbf{q}\cdot\mathbf{r}}+\int\limits_{a<r\leqslant
   b}d^3r\,\frac{a^3}{r^3}\left(\frac{z^2}{r^2}-\frac{1}{3}\right)
   e^{-i\mathbf{q}\cdot\mathbf{r}}\right].\label{eq:appendix_Iz}
\end{eqnarray}
Here, the parameters $V_0$, $a$ and $b$ have been defined in the context.  Due
to the factor $\exp(-i\mathbf{q}\cdot\mathbf{r})$, it is not obvious how to
proceed the integrations in the above equations.  However, if we rotate the coordinate
system so that $\mathbf{q}$ lies on the $z^\prime$-axis, we can take advantage
of the cylindrical symmetry around the $z^\prime$-axis to cancel odd symmetric terms.

Let $\theta_q$ and $\phi_q$ be the polar and azimuth angles of
$q$, i.e.,
\begin{equation}
(\hat{q}_x,\hat{q}_y,\hat{q}_z)=(\sin\theta_q\cos\phi_q,\sin\theta_q\sin\phi_q,\cos\theta_q).
\end{equation}
As illustrated in Fig.~\ref{fig:appendix}, a rotation around the $z$-axis
clockwise by $\frac{\pi}{2}-\phi_q$
would turn the $x$-axis to the $x^\prime$-axis so that the latter is
perpendicular to the $\hat{z}-\hat{q}$ plane.  The $y$-axis would now lie on the
$\hat{z}-\hat{q}$ plane.  Then a rotation around the
$x^\prime$-axis clockwise by $\theta_q$ would turn the $z$-axis to the $\mathbf{q}$
direction.  Consequently, a coordinate transformation follows
\begin{eqnarray}
   x&=&x^\prime\sin\phi_q+y^\prime\cos\theta_q\cos\phi_q+
   z^\prime\sin\theta_q\cos\phi_q,\label{eq:coordinate_transformx}\\
   y&=&-x^\prime\cos\phi_q+y^\prime\cos\theta_q\sin\phi_q+
   z^\prime\sin\theta_q\sin\phi_q,\label{eq:coordinate_transformy}\\
   z&=&-y^\prime\sin\theta_q+z^\prime\cos\theta_q.
   \label{eq:coordinate_transformz}
\end{eqnarray}
Substituting
Eqs.~(\ref{eq:coordinate_transformx})-(\ref{eq:coordinate_transformz})
into Eq.~(\ref{eq:appendix_Ix}), we have 
\begin{equation}
   \mathcal{I}_x=-\frac{V_0a^3}{4\pi E_0\lambdabar_0{}^2}\hat{q}_{x}\hat{q}_{z}
   \int\limits_{a<r^\prime\leqslant b}d^3r^\prime
   e^{iqz^\prime}\frac{-{y^\prime}^2+{z^\prime}^2}{{r^\prime}^5}.
   \label{eq:appendix_Ixp}
\end{equation}
In the above substitution and expansion of $xz$, we have dropped the cross-terms
involving $x^\prime y^\prime$, $x^\prime z^\prime$, and $y^\prime z^\prime$
because they all integrate to 0 due to the cylindrical symmetry around the
$z^\prime$-axis.  Derivations for $\mathcal{I}_y$ and $\mathcal{I}_z$ are
similar.  Further straightforward computations result in the scattered spherical
wave
\begin{eqnarray}
   \Psi_\text{Born}^\text{sca}&=&\lambdabar_0\;\frac{V_0}{E_0}\left(\frac{a}{\lambdabar_0}\right)^3
   \left[F(qa)-F(qb)\right]\frac{e^{ikr}}{r}\nonumber\\
   && \left(
   \begin{array}{c}
   -\left[\left(\hat{q}_x^2+\hat{q}_y^2\right)+\dfrac{2}{3}\dfrac{F(qb)}{F(qa)-F(qb)}
   \right]s_u+\hat{q}_z\left(\hat{q}_x-i\hat{q}_y\right)s_d\\
      \hat{q}_z\left(\hat{q}_x+i\hat{q}_y\right)s_u+
     \left[\left(\hat{q}_x^2+\hat{q}_y^2\right)+\dfrac{2}{3}\dfrac{F(qb)}{F(qa)-F(qb)}\right]s_d
   \end{array}
   \right),
   \label{eq:BornPsiScat-b}
\end{eqnarray}
where
\begin{equation}
   F(\rho)=\frac{\sin\rho}{\rho^3}-\frac{\cos\rho}{\rho^2},
   \label{eq:Frho}
\end{equation}
and $q=\left|\mathbf{q}\right|$.  We immediately notice the limits $F(0)=1/3$
and $F(\infty)=0$.  Because $F(\infty)=0$, the limit of
Eq.~(\ref{eq:BornPsiScat-b}) at $b\to\infty$ is obvious.

\bibliographystyle{elsarticle-num}

\providecommand{\noopsort}[1]{}\providecommand{\singleletter}[1]{#1}%

\end{document}